\documentstyle[psfig,amssymb]{mn}

\title[The stellar content of brightest cluster galaxies]
{The stellar content of brightest cluster galaxies}
  
\author[P.A. James \& B. Mobasher] 
{P.A. James$^1$ and B. Mobasher$^{2,3}$\\  
$^1$ Astrophysics Research Institute, Liverpool 
John Moores University, Egerton Wharf, Birkenhead, CH41 1LD, UK\\
$^2$Astrophysics Group, Blackett Laboratory, Imperial College, 
Prince Consort Rd, London SW7 2BZ, UK\\
$^3$Space Telescope Science Institute, 3700 San Martin Drive,
Baltimore, MD 21218, USA\\
}
\begin{document} 
\maketitle
\begin{abstract}
We present near-infrared K-band spectroscopy of 21 elliptical or cD
Brightest Cluster Galaxies (BCGs), for which we have measured the
strength of the 2.293~$\mu$m CO stellar absorption feature.  We find
that the strength of this feature is remarkably uniform among these
galaxies, with a smaller scatter in equivalent width than for the
normal elliptical population in the field or clusters. The scatter for
BCGs is 0.156~nm, compared with 0.240~nm for Coma cluster ellipticals,
0.337~nm for ellipticals from a variety of other clusters, and
0.422~nm for field ellipticals. We interpret this homogeneity as being
due to a greater age, or more uniform history, of star formation in
BCGs than in other ellipticals; only a small fraction of the scatter
can be due to metallicity variations, even in the BCGs.
Notwithstanding the small scatter, correlations are found between CO
strength and various galaxy properties, including R-band absolute
magnitude, which could improve the precision of these galaxies as
distance indicators in measurements of cosmological parameters and
velocity flows.
\end{abstract}
  
  
\section {Introduction}

Elliptical or cD Brightest Cluster Galaxies (BCGs) have long been used
as cosmological probes, in attempts to determine parameters such as
H$_0$ and q$_0$ (Sandage 1972; Sandage \& Hardy 1973; Hoessel, Gunn \&
Thuan 1980; Sandage \& Tammann 1990). These studies make use of the
ease of selection of such objects at cosmological distances, which is
a result of their bright optical luminosities and privileged locations
defined by other galaxies and by the centroids of cluster X-ray
emission.  They also appeal to another, more surprising property of
BCGs: they appear to be excellent standard candles, with Sandage \&
Hardy \shortcite{sa:73} finding a scatter of only 0.28~mag in visual
absolute magnitude, after correction for cluster Bautz-Morgan
\shortcite{ba:70} and richness class.  Lauer \& Postman
\shortcite{la:94}, in a study using more modern CCD techniques, found a
scatter of 0.33~mag in R-band absolute magnitudes of BCGs.  They were
able to reduce this scatter to 0.25~mag by using the relation of
absolute magnitude with the `Structure Parameter' $\alpha$ of Hoessel
\shortcite{ho:80}, which is defined as
$$
\alpha= \left.\frac{d \log L_m}{d \log r}\right|_{r_m}
$$ 
where $r_m$ is a metric radius of 9.6~kpc and $L_m$ is the luminosity
within that metric radius.  Hoessel \shortcite{ho:80} explains
$\alpha$ as being a dimensionless parametrization of galaxy size, and
empirically it is found to increase with BCG luminosity. Indeed he
found a scatter in $\alpha$-corrected BCG absolute magnitudes of just
0.21~mag, even lower than that of Lauer \& Postman
\shortcite{la:94}.

This uniformity in BCG luminosities remains something of a mystery.
There is indeed good reason for expecting a wider disparity in BCG
properties than in those of other galaxy types.  Hoessel
\shortcite{ho:80} and Lauer \shortcite{la:88} showed that $\sim$30\% of BCGs
have multiple nuclei, a frequency which is greater than would be
predicted by chance superpositions, and which indicates that galactic
cannibalism is very common in BCGs at the present epoch.  The
continued accretion of cluster members would seem to ensure that BCGs
as a population should have diverse properties, and it would seem
probable that this should be reflected both in their total luminosity
and in their composition.  One motivation for the present paper is to 
test whether BCGs are more homogeneous or more diverse than the
general population of elliptical galaxies.

In this paper, we look at the stellar populations of BCGs, by
measuring the strength of the 2.293$\mu$m CO absorption feature for 21
BCGs, and comparing the distribution of values with similar
measurements for a large sample of field, group and cluster
ellipticals (Mobasher \& James 1996; James \& Mobasher 1999; Mobasher
\& James 2000).  CO strength
contains information on recency of star formation, since it is very
strong in supergiants (present 10$^7$--10$^8$ years after a burst of
star formation), strong in the cool AGB stars which contribute
significantly to the near-IR light after 10$^8$--10$^9$ years (Renzini
\& Buzzoni 1986; Oliva et al. 1995), and somewhat weaker in older
populations.  It also displays some metallicity dependence, being weak
in very low metallicity globular clusters \cite{or:97}.  This
dependence was quantified by Doyon, Joseph \& Wright
\shortcite{do:94}, and further studied in Mobasher \& James
\shortcite{mo:99}.

Such studies of BCGs are particularly significant considering the
apparent large-scale velocity flow found by Lauer \& Postman
\shortcite{la:94}. Using a sample of 119 BCGs out to
a redshift of 15000~kms$^{-1}$, they found the restframe defined by
the galaxies to differ from that of the Cosmic Microwave Background by
almost 700~kms$^{-1}$.  This result has been interpreted as evidence
for a cosmological streaming flow, but an alternative explanation
would be that BCG properties vary systematically around the sky, for
example due to stellar population changes from galaxy to galaxy.  This
provides a further motivation for the present study.

The organisation of this paper is as follows.  Section 2 describes the
selection of target galaxies, the observations, and the data
reduction.  Section 3 contains the main results, including a
comparison of CO absorption strengths of BCGs and other elliptical
galaxies, and correlations of CO strengths with other galaxy
parameters.  Section 4 summarises the main conclusions.

\section {Sample Selection, Observations and Data Reduction}
  
The galaxy sample was selected from the BCG list of Lauer \& Postman
\shortcite{la:94}.  All have measured recession velocities less than
15,000~kms$^{-1}$, with R band photometry presented in Lauer \&
Postman \shortcite{la:94}.  The observations presented here were carried
out using the United Kingdom Infrared Telescope (UKIRT) during the 4
nights of 21--24 February 1999. The instrument used was the long-slit
near-IR spectrometer CGS4, with the 40~line~mm$^{-1}$ grating and the
long-focal-length (300~mm) camera. The 4-pixel-wide slit was chosen,
corresponding to a projected width on the sky of 2.4~arcsec. Working
in 1st order at a central wavelength of 2.2 $\mu m$, this gave
coverage of the entire K window. The CO absorption feature, required
for this study, extends from 2.293 $\mu m$ (rest frame) into the
K-band atmospheric cut-off. The principal uncertainty in determining
the absorption depth comes from estimating the level and slope of the
continuum shortward of this absorption which requires wavelength
coverage down to at least 2.2 $\mu m$ and preferably to
shorter wavelengths. There are many regions of the continuum free from
lines even at this relatively low resolution. The effective
resolution, including the degradation caused by the wide slit, is
about 230.

For each observation, the galaxy was centred on the slit by maximising
the IR signal, using an automatic peak-up facility.  Total on-chip
integration times of 12 minutes were used for the brightest and most
centrally-concentrated ellipticals while an integration time of 24
minutes was more typically required. During this time, the galaxy was
slid up and down the slit at one minute intervals by 22~arcsec, giving
two offset spectra which were subtracted to remove most of the sky
emission. Moreover, the array was moved by 1 pixel in the spectral
direction between integrations to enable bad pixel replacement in the
final spectra. Stars of spectral types A0--A6, suitable for monitoring
telluric absorption, were observed in the same way before and after
each galaxy, with airmasses matching those of the galaxy observations
as closely as possible. Flat fields and argon arc spectra were taken
using the CGS4 calibration lamps. A total of 21 brightest cluster
galaxies was observed.

The data reduction was performed using the FIGARO package in the
STARLINK environment. The spectra were flatfielded and polynomials
fitted to estimate and remove the sky background. These spectra
were then shifted to the rest frame of the galaxy, using redshifts
from Lauer \& Postman \shortcite{la:94}. The atmospheric transmissions
were corrected by dividing each spectrum with the spectrum of the star
observed closely in time to the galaxy, and at a similar airmass. The
resulting spectrum was converted into a normalised, rectified spectrum
by fitting a power-law to featureless sections of the continuum and
dividing the whole spectrum by this power-law, extrapolated over the
full wavelength range.  Two rectified spectra are shown in Fig. 1. The
apparent emission features at 2.14~$\mu$m and 2.10~$\mu$m are
artefacts caused by absorptions in the A stars used for atmospheric
transmission correction, and appear in different positions because of
the restframe corrections.

To measure the depth of the CO absorption feature, the procedure
outlined in James \& Mobasher \shortcite{ja:99} is used.  The restframe,
rectified spectra were rebinned to a common wavelength range and
number of pixels, to avoid rounding errors in the effective wavelength
range sampled by a given number of pixels.  The CO strength for each
spectrum was determined using the method of Puxley, Doyon \& Ward
\shortcite{pu:97}.  They advocate the use of an equivalent width,
CO$_{EW}$, which is determined within the CO absorption feature between
rest-frame wavelengths of 2.293~$\mu$m and 2.320~$\mu$m. This
wavelength range was found by Puxley et al. \shortcite{pu:97} to give
maximum sensitivity to stellar population variations, and can be used
for galaxies with recession velocities of up to $\sim$18000~kms$^{-1}$
before the spectral region of interest shifts out of the usable K
window.  This is not the case for the CO index CO$_{sp}$ used by Doyon
et al. \shortcite{do:94}, which extends over a restframe wavelength range of
2.320--2.400~$\mu$m and would have been affected by large and uncertain
telluric absorption and emission for the highest redshift galaxies in the
present sample.  Thus we only present CO$_{EW}$ values in this paper.

A further advantage of the CO$_{EW}$ definition of Puxley et al.
\shortcite{pu:97} is that it is almost completely unaffected by
velocity dispersion effects, due to the wide range of wavelength over
which the absorption is measured.  Puxley et al. \shortcite{pu:97}
find the velocity dispersion corrections to be insignificant, which
we confirmed by smoothing low-velocity-dispersion galaxy spectra to
an effective velocity dispersion of 500~kms$^{-1}$.  The resulting
change in CO$_{EW}$ was $\sim$0.25\%, very much smaller than the
random errors. 

The errors on the CO$_{EW}$ values include three components. The
first was calculated from the standard deviation in the fitted
continuum points, on the assumption that the noise level remains
constant through the CO absorption, giving an error on both the
continuum level and on the mean level in the CO absorption, which were
added in quadrature.  The second error component comes from the formal
error provided by the continuum fitting procedure.  This procedure
could leave a residual tilt or curvature in the spectrum, and the
formal error was used to quantify this contribution.  The final
component was an estimate of the error induced by redshift and
wavelength calibration uncertainties. All three errors were of similar
sizes, with only the first varying from spectrum to spectrum, as a
result of signal-to-noise variations (see Fig. 1), and all three were
added in quadrature to give the value quoted in Table 1.

\begin{figure}
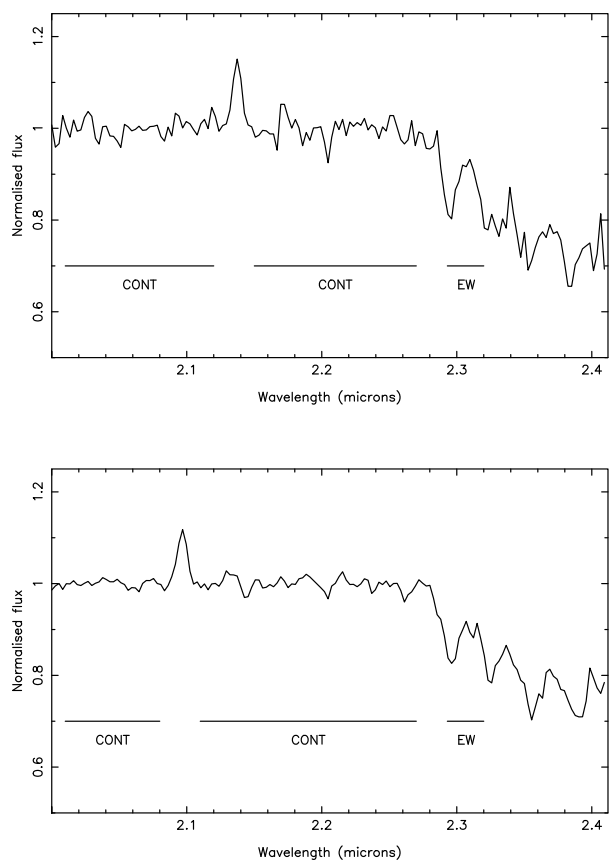
 
\centerline{\psfig{figure=bcg_fig1a.eps,width=0.5\textwidth,angle=270}}
\centerline{\psfig{figure=bcg_fig1b.eps,width=0.5\textwidth,angle=270}}
\caption{The rectified spectra of the BCGs in Abell~1060 (top) and
Abell~1016 (bottom), showing the wavelength ranges used to define the
continuum level and the CO Equivalent Width.  These spectra illustrate
the lowest and highest signal-to-noise spectra used for the present
paper. }
\end{figure}

\section{Results}

The equivalent widths of CO absorption features for the sample of BCGs
observed in this study are presented in Table 1.  The data included in
this table are Abell
\shortcite{ab:58} catalogue numbers (column 1), BCG names (column 2),
CO$_{EW}$ values with 1--$\sigma$ errors (column 3), recession
velocity in kms$^{-1}$ (column 4), absolute R-band magnitude
corresponding to the metric luminosity $L_m$ (column
5), structure parameter ($\alpha$) (column 6) and the magnitude
residual relative to the best-fit $L_m$--$\alpha$ relation (column 7),
(columns 4--7 are all taken from Lauer \& Postman \shortcite{la:94},
who assumed a Hubble constant of 80~kms$^{-1}$Mpc$^{-1}$).  Columns 8
and 9 contain velocity dispersions and Mg$_2$ metallicity indices,
where available, from Faber et al. \shortcite{fa:89}.

\begin{table*}
\centering
\begin{minipage}{140mm}
\caption{Photometric and spectroscopic parameters
for 21 brightest cluster galaxies}
\begin{tabular}{llllcclcl}\\ 
 & & & & & & & &\\
Abell$\#$   & Galaxy name & CO$_{EW}$(nm)& V$_{rec}$ & M$_R$ & $\alpha$ &
dM$_{\alpha}$ & $\sigma$ (kms$^{-1}$) & Mg$_2$ \\

& & & & & & &\\

496  & MCG-02-12-039 &   3.36$\pm$0.20 &  9893 & -22.579 &  0.786 & 0.103 &  274 & --\\
533  &  --           &   3.55$\pm$0.19 & 14365 & -22.397 &  0.500 & 0.012 &  --  & --\\
539  & MCG+01-14-019 &   3.36$\pm$0.18 &  9682 & -22.366 &  0.505 & 0.050 &  309 & --\\
548  & ESO488-G033   &   3.23$\pm$0.18 & 11848 & -22.490 &  0.493 &-0.093 &  --  & --\\
569  & NGC2329       &   3.15$\pm$0.18 &  5724 & -22.357 &  0.472 & 0.003 &  271 & 0.269\\
576  &  --           &   3.42$\pm$0.18 & 12072 & -22.051 &  0.296 &-0.083 &  --  & --\\
634  & UGC4289       &   3.20$\pm$0.18 &  8135 & -22.248 &  0.497 & 0.156 &  241 & --\\
671  & IC2378        &   3.44$\pm$0.19 & 14970 & -22.961 &  0.713 &-0.310 &  313 & --\\
779  & NGC2832       &   3.48$\pm$0.18 &  6867 & -23.011 &  0.596 &-0.465 &  354 & 0.340\\
912  &  --           &   3.23$\pm$0.19 & 13572 & -21.984 &  0.423 & 0.283 &  --  & --\\
957  & UGC5515       &   3.39$\pm$0.19 & 13438 & -22.882 &  0.760 &-0.207 &  350 & --\\
999  & MCG+02-27-004 &   3.34$\pm$0.18 &  9749 & -22.328 &  0.443 &-0.021 &  --  & --\\
1016 & IC613         &   3.28$\pm$0.18 &  9705 & -22.112 &  0.436 & 0.181 &  --  & --\\
1060 & NGC3311       &   3.39$\pm$0.24 &  3704 & -22.392 &  0.845 & 0.297 &  192 & 0.297\\
1142 & IC664         &   3.34$\pm$0.18 & 10118 & -22.367 &  0.544 & 0.110 &  --  & --\\
1656 & NGC4889       &   3.55$\pm$0.18 &  6497 & -23.106 &  0.612 &-0.541 &  404 & 0.359\\
2147 & UGC10143      &   3.01$\pm$0.18 & 10384 & -22.374 &  0.666 & 0.244 &  303 & --\\
2162 & NGC6086       &   3.25$\pm$0.18 &  9547 & -22.594 &  0.503 &-0.179 &  325 & 0.344\\
2197 & NGC6173       &   3.62$\pm$0.25 &  8800 & -22.988 &  0.592 &-0.466 &  295 & 0.332\\
2199 & NGC6166       &   3.23$\pm$0.25 &  9348 & -22.748 &  0.777 &-0.067 &  320 & 0.340\\
2634 & NGC7720       &   3.60$\pm$0.25 &  9141 & -22.662 &  0.643 &-0.065 &  305 & 0.339\\

\end{tabular}
\end{minipage}
\end{table*}

\begin{figure}
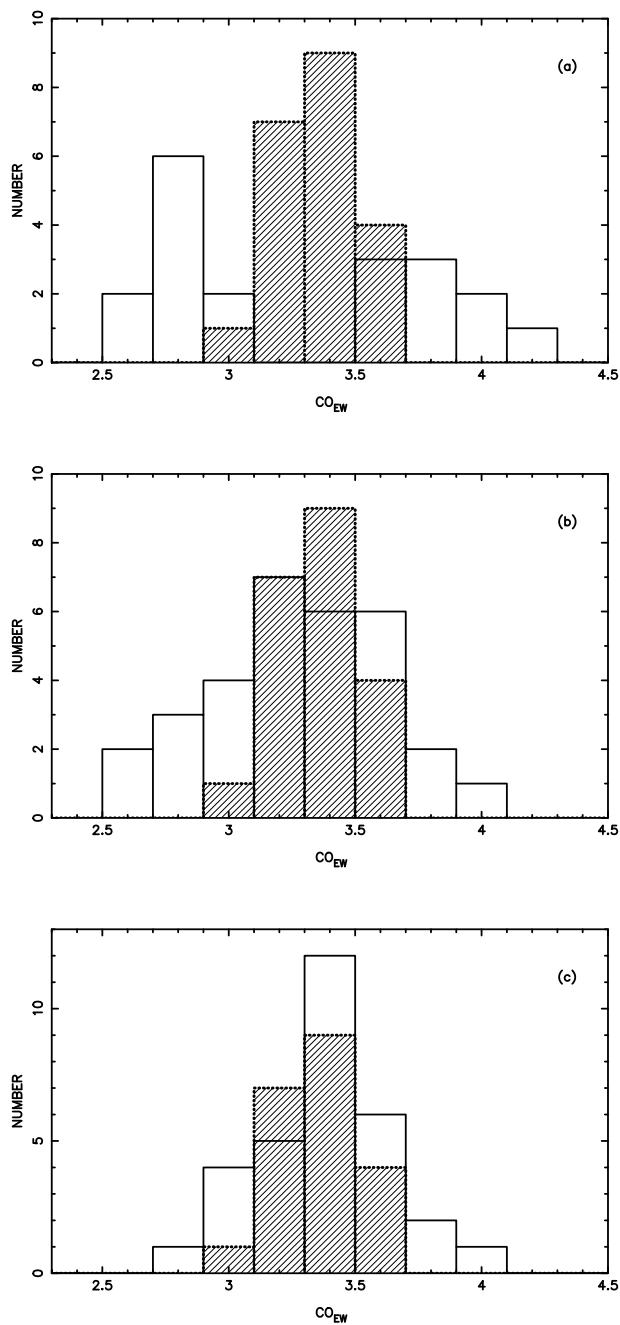
 
\centerline{\psfig{figure=bcg_fig2a.eps,width=0.5\textwidth,angle=270}}
\centerline{\psfig{figure=bcg_fig2b.eps,width=0.5\textwidth,angle=270}}
\centerline{\psfig{figure=bcg_fig2c.eps,width=0.5\textwidth,angle=270}}
\caption{Histograms of CO$_{EW}$ distributions BCGs (hashed regions), overlaid 
on field and group galaxies (a), cluster galaxies (b), and Coma 
cluster galaxies (c).}
\end{figure}

We find the mean CO$_{EW}$ value for the 21 BCGs (3.35$\pm$0.03) to be
effectively identical to that of the Coma cluster ellipticals
(3.37$\pm$0.04) \cite{mo:99}, and to that of the 31 ellipticals from a
range of clusters discussed by James \& Mobasher \shortcite{ja:99}
(3.29$\pm$0.06).  The cluster and BCG distributions lie between
the distributions of `isolated' and `group' field ellipticals discussed by
James \& Mobasher
\shortcite{ja:99} and shown in Fig. 2a.  The major difference between the
BCG CO$_{EW}$ values and those of other ellipticals is the remarkably
small range in the former: the standard deviation for BCGs is 0.156,
compared to 0.240 for Coma ellipticals, 0.337 for general cluster
galaxies, and 0.422 for cluster plus field ellipticals.  Indeed, the
scatter in BCG CO absorption strengths is that predicted from the
error estimates on the individual CO$_{EW}$ values, and so the
intrinsic scatter may be much smaller still.  Given the small number
of BCG galaxies, a Kolmogorov-Smirnov test cannot distinguish between
the distributions of BCG and cluster or Coma galaxies in Figs. 2b and
2c, but there is less than 10\% chance that the BCGs are drawn from
the same parent population as all the non-BCG ellipticals, and less
than 1\% chance that they are from the same population as field
ellipticals (Fig. 2a).

The BCGs are drawn from a much narrower region of the galaxy
luminosity function than are the comparison samples in Fig. 2, which
could affect the interpretation of this result.  The 21 BCGs have a
range in M$_R$ of -22.0 to -23.1, little more than a magnitude. R-band
photometry is not available for all the comparison galaxies, but good
estimates can be made from published optical and near-IR photometry,
leading to an estimated range of M$_R$ of -19.4 to -22.8 for the Coma
cluster ellipticals, and -19.8 to -22.5 for the field and cluster
ellipticals discussed by James and Mobasher
\shortcite{ja:99}.  We investigated whether the differences in
CO$_{EW}$ scatter shown in Fig. 2 result from these differences in
luminosity range by regressing CO$_{EW}$ on absolute magnitude, and
studying the distributions of CO$_{EW}$ residuals about the best-fit
lines.  The distributions of these residuals are shown in Fig. 3.  The
dashed, diagonally shaded columns represent the residuals for the
BCGs; the thick, dotted columns are those for the Coma cluster
galaxies; and the solid lines represent the residuals for the field,
group and cluster galaxies from James and Mobasher \shortcite{ja:99}.
The standard deviations of the CO$_{EW}$ residuals are 0.133~nm for
the BCGs, 0.221~nm for the Coma ellipticals, and 0.419~nm for the
cluster plus field ellipticals. This reinforces the conclusion from
Fig. 2 that the BCGs have substantially more homogeneous CO strengths
than the other elliptical galaxies studied, and this result does not
appear to be a selection effect caused by the small luminosity range
of the BCGs.

\begin{figure} 
\centerline{\psfig{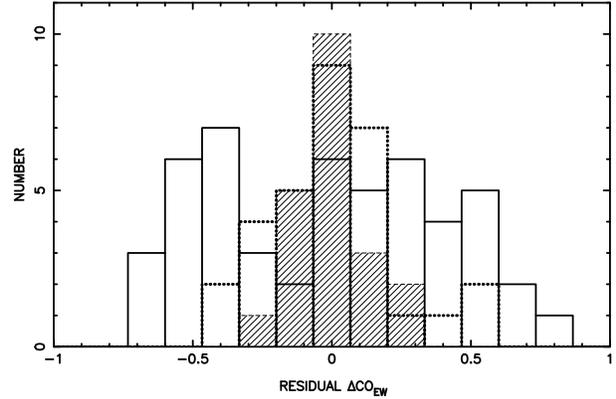}}
\caption{Histograms of residuals of CO$_{EW}$ when regressed on
absolute magnitude, for BCGs (hashed regions with dashed lines),
overlaid on residuals for Coma ellipticals (thick dotted lines) and
general field and cluster ellipticals (solid lines).}
\end{figure}

This uniformity in CO$_{EW}$ values is the main result of this paper,
and it is important to consider what it implies in terms of
differences between BCGs and other ellipticals.  Both high metallicity
and recency of star formation are expected to increase CO$_{EW}$
values. The effect of metallicity on CO$_{EW}$ values can be estimated
for the galaxies with measured Mg$_2$ indices, using the following
method.  From Fig. 37 of
Worthey \shortcite{wo:94}, a change in [Fe/H] from -0.25 to 0.00
changes the Mg$_2$ index from 0.216 to 0.258, and the change is
approximately linear over the modelled range.  Thus we infer a
relation of the form 
$$\delta Mg_2 = 0.168~\delta [Fe/H].$$  
Doyon et al.
\shortcite{do:94} find a relation between [Fe/H] and their CO index
CO$_{sp}$, $$\delta CO_{sp} = 0.11~\delta [Fe/H],$$ and from the
definitions in Puxley et al. \shortcite{pu:97} it is straightforward
to convert from the index CO$_{sp}$ to CO$_{EW}$. Then, the measured
scatter in Mg$_2$ index of 0.029 for the BCGs should cause a scatter
of 0.060~nm in CO$_{EW}$, 38\% of the observed scatter.  For Coma
ellipticals, the measured Mg$_2$ scatter is 0.024, equivalent to a
scatter of 0.049~nm in CO$_{EW}$, 20\% of that observed, and for the
field and cluster sample, the Mg$_2$ scatter is 0.030, and the
predicted CO$_{EW}$ scatter 0.062~nm, 15\% of that observed. Note also
that the scatters in Mg$_2$ values are very similar in the three
subsamples, whereas they have very different CO$_{EW}$ distributions.
Thus, we conclude that metallicity differences have little effect on
the measured CO$_{EW}$ values for the elliptical galaxies studied
here, and propose that star formation history is the dominant factor
causing the larger scatter for non-BCG ellipticals. If so, the
differences in the distributions of CO$_{EW}$, shown in Fig. 2 would
be the result of wider variations in star formation history for
general field and cluster ellipticals than for the BCGs. This
indicates that BCGs formed their stars very early; if there has been
more recent star formation in these galaxies then the rate of star
formation as a function of epoch must have been very uniform from
galaxy to galaxy.

Given the narrow range in BCG CO$_{EW}$ values, it is unrealistic to
expect very strong correlations with other BCG parameters.
Nevertheless, Fig. 4 does show a good correlation with absolute R-band
magnitude in a 10~kpc metric aperture, M$_R$, with a correlation
coefficient of 0.51 and a probability of 98.4\% that this represents a
true correlation (i.e. 1.6\% probability that it could arise by
chance). This is significant enough to be useful as a distance
indicator: the scatter in M$_R$ for the 21 galaxies observed is
0.326~mag, which reduces to 0.280~mag when the M$_R$ values are
corrected for the CO$_{EW}$ effect. The slope of the regression line
of M$_R$ on CO$_{EW}$ is somewhat smaller than that for the trend in
M$_K$ (total K-band absolute magnitude) vs CO$_{EW}$ for Coma cluster
ellipticals \cite{mo:99}, at -1.1$\pm$0.5~mag/nm c.f. -1.6$\pm$0.7~mag/nm for the 
Coma galaxies.  However, this difference is not statistically
significant ($\sim$0.6$\sigma$).  It is not possible to determine
whether the absolute magnitude--CO$_{EW}$ relations are consistent for
the various samples because of the lack of homogeneous photometry, and
the consequent need for large and uncertain colour and aperture
corrections.

\begin{figure} 
\centerline{\psfig{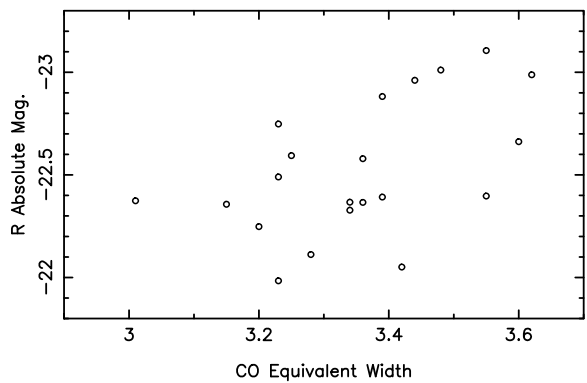}}
\caption{Absolute R-band magnitude versus CO Equivalent Width for the
21 Brightest Cluster Galaxies. }
\end{figure}

Similarly, there is a strong correlation between CO$_{EW}$ and
residuals (dM$_{\alpha}$) about the relation of M$_R$ with structure
parameter $\alpha$
\cite{ho:80} (Fig. 5), in the sense that galaxies with high CO$_{EW}$ tend
to be bright relative to the mean relation (correlation coefficient
0.60, significance 99.6\%). The residuals (dM$_{\alpha}$) are reduced
from 0.243~mag to 0.195~mag by correcting for the trend with CO$_{EW}$
shown in Fig. 5.  There is no correlation between CO$_{EW}$ and the
structure parameter $\alpha$ itself.

\begin{figure} 
\centerline{\psfig{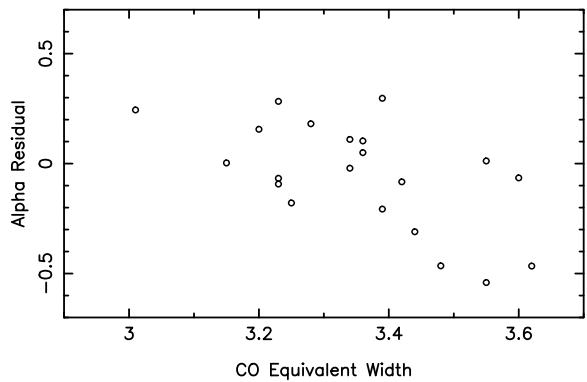}}
\caption{Structure parameter residuals versus CO Equivalent Width for the
21 Brightest Cluster Galaxies. }
\end{figure}

Given the trends found in Figs. 4 \& 5, it is instructive to explore
if these effects could cause the putative streaming flow signal
detected by Lauer
\& Postman \shortcite{la:94} using the full sample of BCGs. However,
we find no significant correlation between CO$_{EW}$ and direction on
the sky (Fig. 6).  This implies that the Lauer \& Postman
\shortcite{la:94} apparent detection of a bulk flow was not an
artefact of differing stellar populations between sample galaxies,
although we have of course only looked at a small fraction (18\%) of their
sample.

\begin{figure} 
\centerline{\psfig{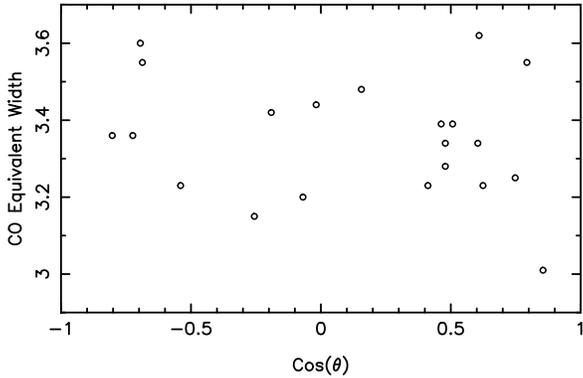}}
\caption{CO$_{EW}$ versus the cosine of the angle between
galaxy direction and the apex of the Lauer \&
Postman\shortcite{la:94} streaming motion, for the
21 Brightest Cluster Galaxies. }
\end{figure}

The trends in figures 4 \& 5 are both in the sense that brighter
galaxies have higher indices: that in Fig. 4 could be a consequence of
a metallicity--absolute magnitude relation, and there is indeed
evidence of a weak correlation of CO$_{EW}$ with metallicity (Fig. 7).
The correlation coefficient here is 0.52, but the relation
has only 80\% significance due to only 8 BCGs having tabulated Mg$_2$
values. Fig. 7 also shows the trend in CO$_{EW}$ with metallicity for
31 Coma cluster galaxies \cite{mo:99}. The BCGs clearly lie at higher
mean metallicity than do the Coma cluster ellipticals (mean Mg$_2$
values 0.328 for the BCGs and 0.294 for the Coma ellipticals).  Using
the relation between Mg$_2$ and [Fe/H] from Worthey
\shortcite{wo:94}, and that between [Fe/H] and CO absorption strength
found by Doyon et al. \shortcite{do:94}, this predicts a difference in
mean CO$_{EW}$ of 0.05~nm, compared to the observed difference of
0.04$\pm$0.07~nm in the same sense. The small variation found here
confirms the weakness of the dependence of CO$_{EW}$ on metallicity,
as found by Doyon et al. \shortcite{do:94}, at least at the high
metallicity values typical of centres of bright galaxies, and also
confirms our earlier conclusion that star formation history is the
dominant effect in determining CO$_{EW}$ strength. This is further
confirmed by the wide spread in the CO$_{EW}$ values of isolated and
group ellipticals \cite{ja:99}, with no corresponding change in their
metallicity.

Finally, we find a weak correlation of CO$_{EW}$ with velocity
dispersion for 14 galaxies with data in Table 1 (Fig. 8). This
corresponds to a correlation coefficient of 0.294, and is significant
at the 69\% level. The slope and correlation coefficient are the same
as was found for 31 Coma cluster ellipticals, also plotted in Fig. 8,
but the mean correlation for the BCGs is again offset, to higher
velocity dispersion at a given CO$_{EW}$. The mean CO$_{EW}$ is almost
identical for the two samples, whilst the BCGs have a much higher
average velocity dispersion, and hence mass, as expected.

\begin{figure} 
\centerline{\psfig{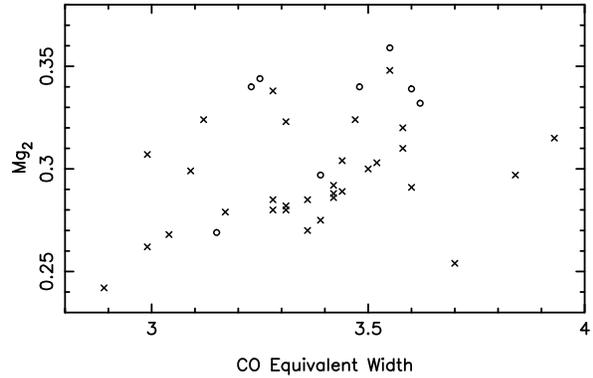}}
\caption{Mg$_2$ index versus CO Equivalent Width for 
the Brightest Cluster Galaxies (circles) and Coma cluster ellipticals
(crosses). }
\end{figure}

\begin{figure} 
\centerline{\psfig{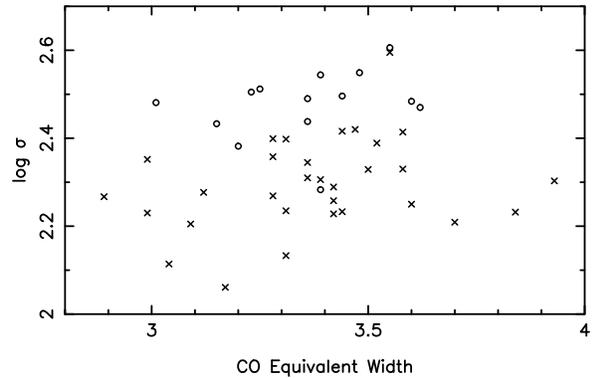}}
\caption{Log of velocity dispersion (kms$^{-1}$) versus CO Equivalent Width for 
the Brightest Cluster Galaxies (circles) and Coma cluster ellipticals
(crosses). }
\end{figure}

\section{Conclusions}

We find that BCGs are much more homogeneous in evolved red stellar
content than ellipticals overall, and BCGs are somewhat more homogeneous
than Coma cluster ellipticals.  The measured scatter in the CO$_{EW}$
indices for BCGs is comparable to the measurement errors.  We
interpret this as implying a more uniform and probably earlier star
formation history for BCGs than for normal ellipticals.  Metallicity
does not appear to be the controlling parameter of CO absorption
strength for elliptical galaxies.

Absolute magnitudes, and magnitude residuals relative to the structure
parameter relation of Hoessel \shortcite{ho:80} correlate well with CO
absorption depth.  This may imply the presence of an additional
intermediate-age population, or a higher metallicity population, in
the galaxies which are over-luminous relative to the mean relation.
However, this effect is very small and such populations, if present,
must be weaker than in most field and cluster ellipticals, given the
high degree of homogeneity in BCG CO$_{EW}$ values.

It may be possible to use these correlations to define higher
precision distance indicators for BCGs, which could be used for
galaxies with redshifts up to 18,000~kms$^{-1}$, as is indicated by
the reduced scatter in R-band absolute magnitude, and the reduced
scatter about the structure parameter relation after correction for
the correlation with CO$_{EW}$, discussed in section
3.

Recent ROSAT observations reveal that a significant number of the
Lauer \& Postman
\shortcite{la:94} galaxies do not lie at the X-ray centroids of
their clusters (Paul
Lynam, private communication), and there may be better candidates for
the dominant central cluster galaxy. Thus our
conclusions may refer more generally to bright galaxies towards cluster
centres than to individual BCGs inhabiting the very centre of the
cluster potential.

\section{Acknowledgements}

We acknowledge the anonymous referee for many useful suggestions.
The United Kingdom Infrared Telescope is operated by the Joint
Astronomy Centre on behalf of the UK Particle Physics and Astronomy
Research Council.

\end{document}